          [2001/04/25 1.1 (PWD)]
\documentclass[a4paper,twocolumn]{esapub} 
\usepackage{natbib}
\usepackage{graphicx}

\title{A submillimeter line survey of low-mass protostars: prelude to ALMA and
Herschel}
\author{E.F. van Dishoeck}
\author{J.K. J{\o}rgensen}
\affil{Leiden Observatory, The Netherlands}
\author{S. Maret}
\author{C. Ceccarelli} 
\affil{Observatoire de Grenoble, France}
\author{E. Caux}
\affil{CESR, Toulouse, France}
\author{F.L.\ Sch\"oier}
\affil{Stockholm Observatory, Sweden}
\author{A.\ Castets}
\affil{Observatoire de Bordeaux, France}
\author{A.G.G.M. Tielens}
\affil{Kapteyn Astronomical Institute, Groningen, The Netherlands}

\begin{document}


\maketitle

\begin{abstract}
The results from a single-dish molecular line survey of a set of 18
deeply embedded young stellar objects are summarized. More than 40
lines from 16 different species were observed with the JCMT, Onsala,
IRAM 30m and SEST telescopes.  The multi-transition data are analyzed
using a temperature and density structure derived from models of the
dust continuum emission. For the outer envelope ($>$300 AU), the data
indicate a `drop' abundance profile for many species, with normal
abundances in the outer- and innermost regions and highly depleted
abundances in an intermediate zone.  This zone is bounded at the outer
edge by the density where the timescale for freeze-out becomes longer
than the lifetime of the core, and at the inner edge by the
evaporation temperature of the species involved. In the innermost
envelope ($<$300 AU), all ices evaporate resulting in jumps in the
abundances of complex organic molecules such as CH$_3$OH. A key
project for Herschel will be to survey gas-phase water in these
objects, whose abundance shows extreme variations with
temperature. ALMA wil be able to directly image the chemical
variations throughout the envelope and zoom in on the inner hot core
and protoplanetary disks on scales on tens of AU.
\end{abstract}

\section{Introduction}

Low mass stars like our Sun are born deep inside molecular clouds. In
the earliest, deeply-embedded stage ($A_V>100$ mag) the protostar is
surrounded by a collapsing envelope and a circumstellar disk through
which material is accreted onto the growing star. This so-called
`class 0' phase (Andr\'e et al.\ 2000) lasts only a very short time, a
few$\times 10^4$ yr, but is critical for the subsequent evolution of
the star: it is the phase in which the final mass of the star, the
final mass of the circumstellar disk (and thus its ability to form
planets), and the initial chemical composition of the disk are
determined.  Protostars differ from the pre-stellar cores studied
extensively by other groups (e.g., Bergin et al.\ 2001, Tafalla et
al.\ 2002, Lee et al.\ 2003) in that there is a central source which
heats the envelope from the inside and dominates the energy balance,
rather than only weak heating from the outside by the external
radiation field.

To constrain their physical and chemical structure, our groups started
a large single-dish survey of low-mass protostars using the JCMT, IRAM
30m, Onsala and SEST telescopes. Previous studies focused either on
selected objects (e.g., IRAS 16293 -2422: Blake et al.\ 1994, van
Dishoeck et al.\ 1995, Ceccarelli et al.\ 1998, 2000a,b, Sch\"oier et
al.\ 2002; Serpens: Hogerheijde et al.\ 1999), or on specific
molecules such as sulfur-bearing species and deuterated molecules
(e.g., Buckle \& Fuller 2002, 2003, Wakelam et al.\ 2004, Roberts et
al.\ 2002, Loinard et al.\ 2001, Shah \& Wootten 2001). Two important
developments make such a survey timely. The first is the advent of
bolometer arrays to map the continuum emission from cold dust. Model
fits of these maps together with the spectral energy distributions
allow accurate constraints of the dust temperature and density
throughout the envelope, independent from the line observations (e.g.,
Motte \& Andr\'e 2001, Shirley et al.\ 2002, J{\o}rgensen et al.\
2002). The second development is the improved sensitivity of
(sub)millimeter receivers, which makes multi-line observations of even
minor species possible. Combined with well-tested radiative transfer
methods (see van Zadelhoff et al.\ 2002 for overview), measurements of
lines originating from levels with a range in energy and critical
density can be used to constrain not just the molecular abundances,
but even abundance {\it profiles} throughout the envelope.

Our sample consists of 18 objects, containing 12 Class 0 sources taken
from Andr\'e et al.\ (2000) and augmented by 5 Class I objects and 2
pre-stellar cores. The main selection criteria are 
luminosities less than 50 $L_{\odot}$, distances less than 450 pc, and
visible from the JCMT. The distinction between Class 0 and I
objects is somewhat arbitrary and it is even unclear whether these two
sets of objects represent truly different evolutionary stages. In
practice, they represent a division in envelope mass at $\sim$0.5
$M_{\odot}$.

Some of the questions addressed with this data set are: (1) What is
the physical and chemical structure of low-mass protostellar envelopes
down to scales of a few hundred AU? (2) Can we trace the evolution of
low-mass protostars through chemical diagnostics and put constraints
on the timescales?  (3) What is the role of freeze-out and grain
chemistry, and does it differ from that found in high-mass
protostellar regions?  (4) What is the dominant mechanism for
returning molecules from the grains: gentle evaporation due to
radiative heating or violent disruption of grain mantles in shocks?
(5) Do low-mass protostars develop a `hot core' chemistry with
abundant complex organic molecules, as do high-mass protostars?

\section{Observations and analysis}

Most of the data were obtained with the JCMT in 2001--2003 in the 230
and 345 GHz atmospheric windows. These data were complemented by lower
frequency observations using the Onsala, IRAM 30m or SEST
telescopes. In total more than 40 lines of 16 species (CO, CS,
HCO$^+$, DCO$^+$, N$_2$H$^+$, HCN, HNC, DCN, CN, HC$_3$N, SO, SO$_2$,
SiO, H$_2$CO, CH$_3$OH, CH$_3$CN and isotopes) were obtained.  Typical
integration times at the JCMT ranged from 30 to 60 minutes, resulting
in rms noise levels of 30--50 mK in a $\sim$0.5 km s$^{-1}$ velocity
bin. The JCMT beam ranges from 15$''$ to 20$''$, corresponding to
linear scales of a few thousand AU at the typical distances of our
sources.  The observing beams at lower frequencies are up to a factor
of 2 larger. Thus, the envelopes are unresolved in the single-dish
beams and all radial variations are inferred from the analysis of the
multi-transition data.

\begin{figure}
\centering
\includegraphics[width=0.95\linewidth]{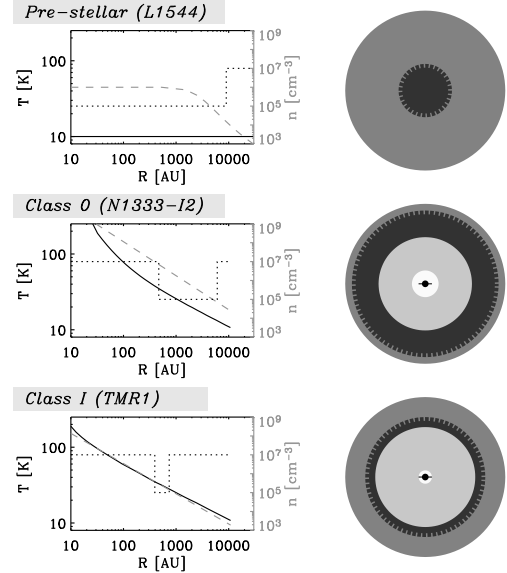}
\caption{Left: density (long-dashed), temperature (solid) 
and abundance (short-dashed) profiles for pre- and protostellar
objects. Right: depletion signature for each
class of object with, from the outside to the inside, the dark grey
indicating the region where the density is too low for significant
freeze-out, the black indicating the region where the molecules are
heavily frozen out, and the light grey indicating the region where
they evaporate (J{\o}rgensen et al.\ 2005a). \label{scenario}}
\end{figure}

For a few objects, interferometer data were obtained with the OVRO,
BIMA and IRAM Plateau de Bure arrays, providing information on scales
of a few hundred AU. This allows the radial structure inferred from
the single-dish data to be tested and may reveal the inner hot cores
directly.

The analysis starts by modeling the dust continuum radiation observed
by SCUBA. Both the radial profile of the emission and the spectral
energy distributions are fitted, assuming spherical symmetry, a
power-law density profile and a dust emissivity which is constant
throughout the envelope. The resulting parameters are the power-law
index of the density structure ($n\propto r^{-p}$, with $p$ typically
1.5--2.0, consistent with collapse models), the ratio of inner and
outer radii, and the optical depth at 100 $\mu$m (J{\o}rgensen et al.\
2002). The dust temperature is calculated by solving the radiative
transfer through the model envelope given the total luminosity of the
source.

The line data are modeled by taking the dust temperature and density
structure as a starting point. A constant gas/dust mass ratio of 100
is assumed and the gas temperature is taken to be equal to that of the
dust.  The molecular excitation and line radiative transfer are then
calculated for each position in the model envelope and the resulting
sky brightness distribution is convolved with the beam profile, or, in
the case of interferometer data, analyzed with the same spatial
filtering. A trial abundance of the molecule under consideration is chosen and
is adjusted until the best agreement with observational data is
reached, e.g., through a $\chi^2$ test. In its simplest form, this
abundance is chosen to be constant with radius, but other abundance
profiles can be tested as well (see \S 3).  See Figure 1 of Doty et
al.\ (2004) for a summary of the modeling procedures. Basic molecular
data such as Einstein-$A$ coefficients and collisional rate
coefficients form an essential input to these models (see Sch\"oier et
al.\ 2005 for summary).

\section{Cold outer envelope}

An inventory of the abundances of various species in the outer
envelope has been presented by J{\o}rgensen et al.\ (2002, 2004a).
The chemical structure of many species follows that of CO.
Assuming a constant abundance, the best-fit results show a clear trend
of increasing CO abundance with decreasing envelope mass.  Objects with
the most massive envelopes have CO abundances that are an order of
magnitude lower than those found for less embedded objects, in which
the abundances approach those found for general molecular clouds.  The
most likely explanation is freeze-out of molecules on icy grains in
the coldest and densest part of the envelope.  Indeed, a so-called
`drop' abundance profile provides a much better fit to the isotopic CO
1--0, 2--1 and 3--2 data than a constant abundance profile.

\begin{figure}
\centering
\includegraphics[width=0.8\linewidth]{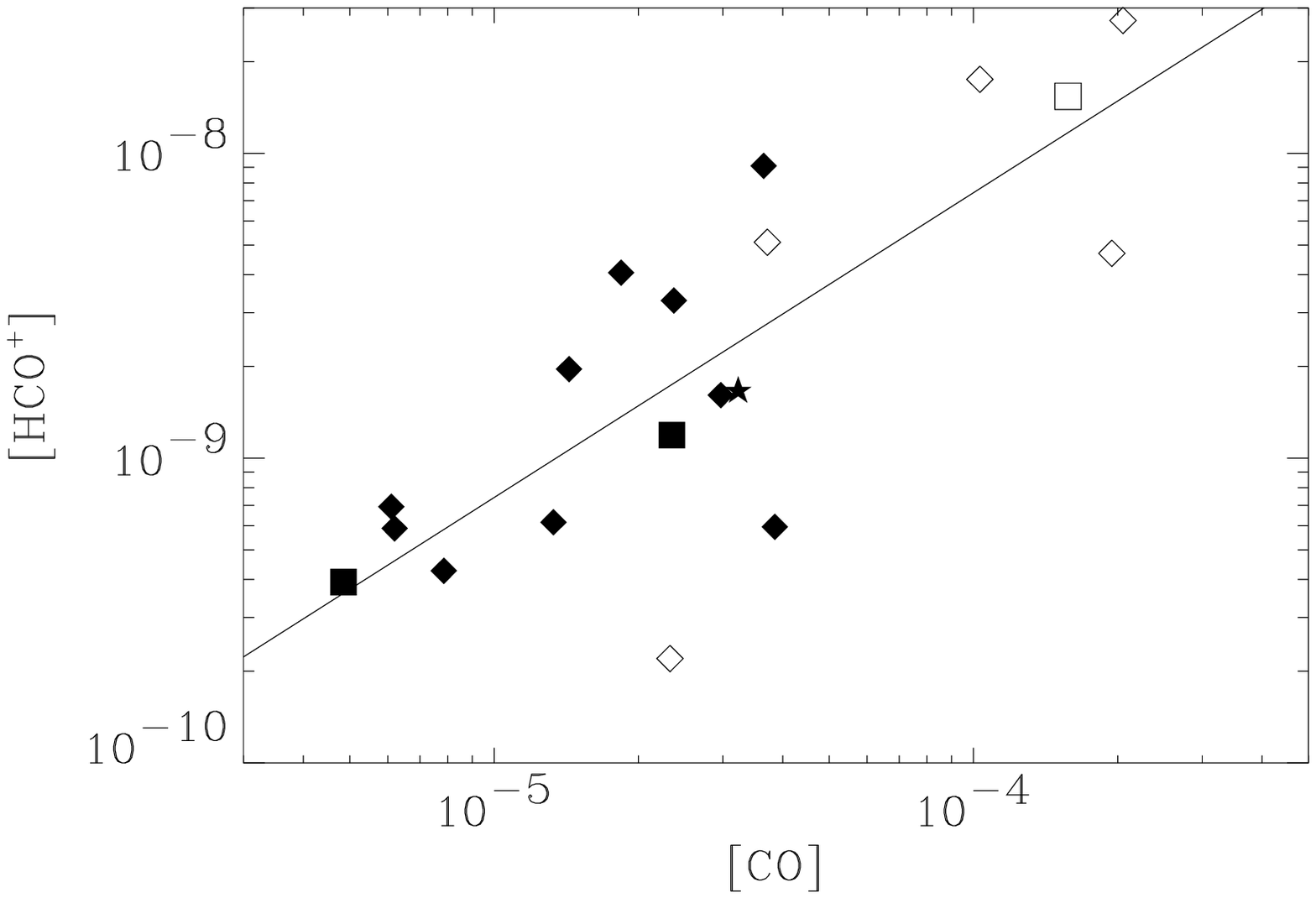}
\includegraphics[width=0.8\linewidth]{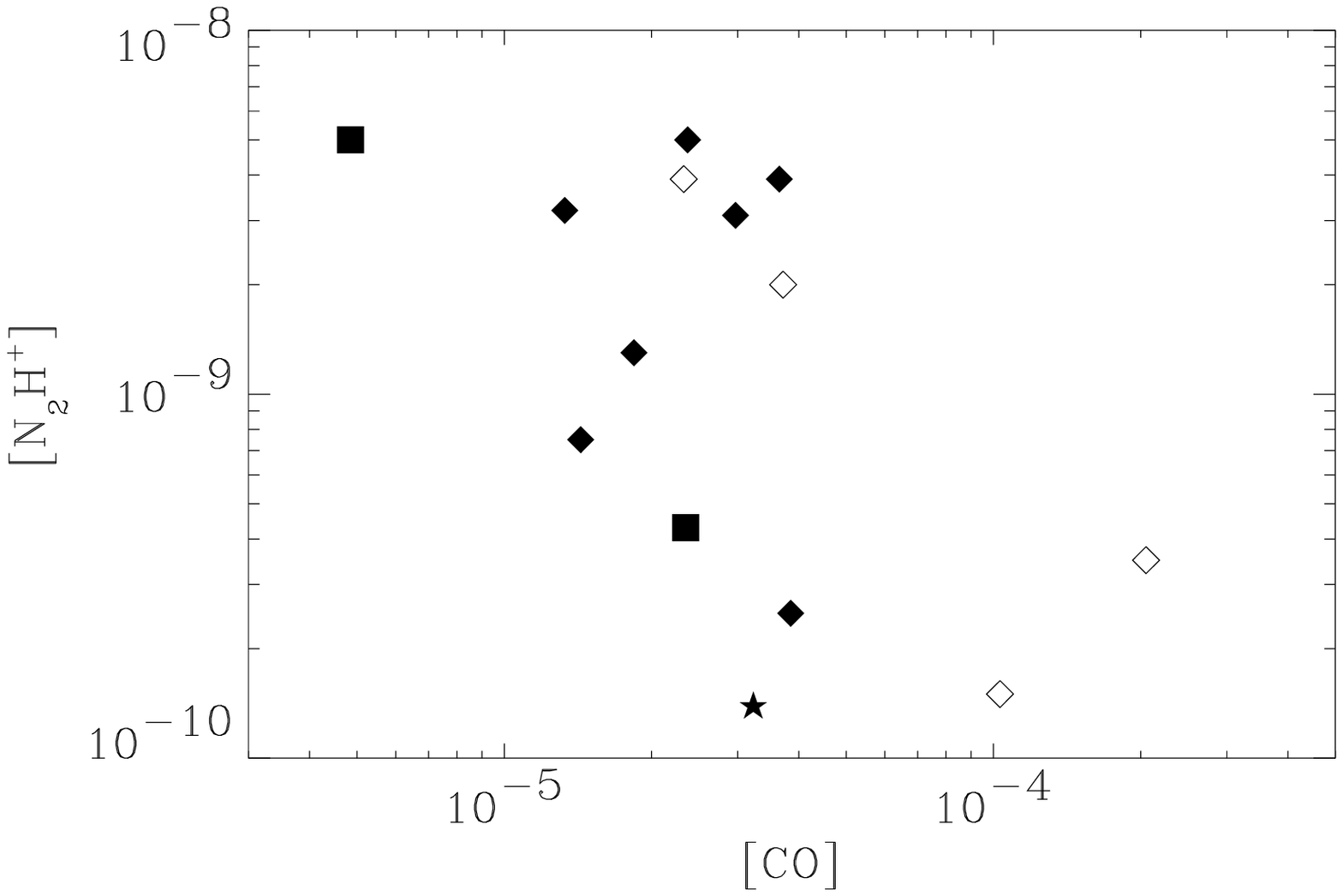}
\caption{The HCO$^+$ (top) and N$_2$H$^+$ (bottom) abudance vs.\ CO 
abundance. Class 0 objects are denoted by filled diamonds; Class I objects
by open diamonds, and pre-stellar cores by filled squares. The Class 0
objects VLA 1623 and IRAS 16293 -2422 are singled out by an open square and
star, respectively (J{\o}rgensen et al.\ 2004a).\label{hco+-n2h+}}
\end{figure}

Figure~\ref{scenario} summarizes the inferred physical and chemical
structure of the pre-stellar cores, Class 0 ($M_{\rm env}>0.5
M_{\odot}$) and Class I ($M_{\rm env}<0.5 M_{\odot}$) objects
(J{\o}rgensen et al.\ 2005a). The left panels show the typical
temperature and density profiles for each type of object. In the
pre-stellar cores, the temperature is low throughout the source and
strong depletion of molecules has been inferred toward the center, but
not the edge, of the core (e.g., Bergin et al.\ 2001, Tafalla et al.\
2002, Lee et al.\ 2003). In the protostellar objects, the central
source sets up a temperature gradient in the envelope, which results
in evaporation of ices in the inner region. CO is one of the most
volatile species and starts to evaporate at $\sim$20 K (Collings et
al.\ 2003). In the outermost region, the density has decreased to the
point where the timescale for freeze-out becomes longer than the age
of the source, resulting in little or no depletion.

The drop abundance profile has four parameters: the outer radius of the
drop zone corresponding to the density $n_{\rm de}$ where depletion
starts; the inner radius of the drop zone where the temperature is
above the evaporation temperature $T_{\rm ev}$; the undepleted
abundance $X_0$ and the depleted abundance $X_D$.  Using the multi-line
data for all sources, the best-fit parameters for CO are $T_{\rm
ev}\approx$35 K, $X_0\approx 2 \times 10^{-4}$, and $X_D$ typically an
order of magnitude lower than $X_0$. The inferred $n_{\rm de}$ is of
order $10^5$ cm$^{-3}$ but varies considerably from object to
object. Thus, the main difference between the Class 0 and I sources is
not the absolute abundance, but rather the width of the depletion
zone, which is significantly smaller for the Class I objects and even
vanishes for objects with $M_{\rm env}<0.1 M_{\odot}$.  If $n_{\rm
de}$ can indeed be related to the age of the core, the inferred values
indicate ages of only $\sim 10^5$ yr, with no significant difference
between the Class 0 and I objects. This suggests that the phase with
heavy depletions (both in the pre- and protostellar stages) is
short-lived, consistent with recent chemical-dynamical models of Lee
et al.\ (2004).

\begin{figure}[h]
\centering
\includegraphics[width=0.8\linewidth]{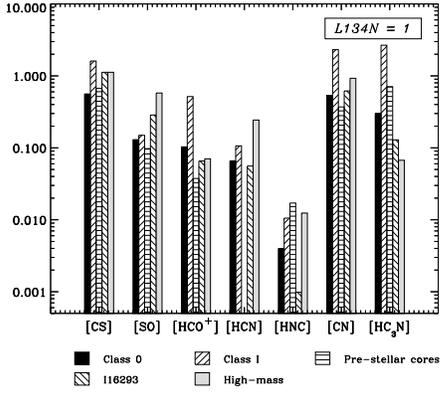}
\caption{Comparison between the average abundances in the outer envelopes
of low-mass Class 0 and I objects, pre-stellar cores, and high-mass
objects. The dark cloud L134N is taken as the reference (J{\o}rgensen
et al.\ 2004a). \label{abundances}}
\end{figure}

The abundance structure of CO is reflected in that of several other
species.  For example, the HCO$^+$ abundance is strongly correlated
with that of CO. Other species such as N$_2$H$^+$ show a clear
anticorrelation because CO is their main destroyer (see
Figure~\ref{hco+-n2h+}).  From the correlation coefficients between
various species, an empirical chemical network can be constructed.
The CS and SO abundances are clearly correlated, as are the CN, HNC
and HC$_3$N abudances. The largest enigma is formed by HCN, whose
abundance does not show any correlation with those of other species,
nor with envelope mass. One complication is that the optically thick
main isotopic HCN lines are often affected by outflows, whereas only
few optically thin lines from minor isotopes have been
detected. Figure~\ref{abundances} compares the inferred abundances in the
low-mass pre- and protostellar objects with those in high-mass
protostars.

The freeze-out of CO is also reflected in the level of deuterium
fractionation.  The most massive Class 0 objects with large freeze-out
zones show very high deuterium fractionation ratios, not just of
DCO$^+$ (Shah \& Wootten 2001, J{\o}rgensen et al.\ 2004a) but even of
doubly deuterated species like D$_2$CO (Loinard et al.\ 2001).  The
deuterium fractionation in the gas phase is initiated by deuteron
transfer from H$_2$D$^+$, whose abundance relative to H$_3^+$ is
enhanced at low temperatures because of its smaller zero-point energy.
In clouds with normal abundances, CO is the prime destroyer of H$_3^+$
and H$_2$D$^+$, but when CO is heavily depleted onto grains, HD
becomes the main reactant of H$_3^+$, enhancing H$_2$D$^+$ even
further (Roberts et al.\ 2003).

\begin{figure}
\centering
\includegraphics[width=0.9\linewidth]{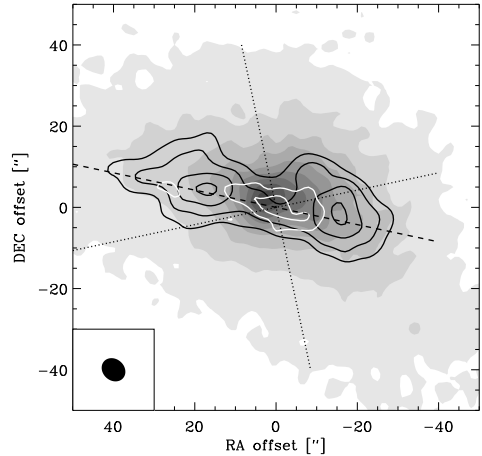}
\caption{Comparison between the N$_2$H$^+$ (black contours) and 
C$^{18}$O (white contours) emission in L~483 obtained with the Owens
Valley Millimeter Array. The grey-scale image indicates the JCMT-SCUBA
450 $\mu$m data. Note the anticorrelation between CO and N$_2$H$^+$
(J{\o}rgensen 2004).\label{l483}}
\end{figure}

The physical and chemical structure presented in Figure 1 has been
confirmed by interferometer studies at higher angular resolution for a
selected set of objects. First, analysis of the continuum data has
shown that the power-law density structure derived from single-dish
data on scales of a few thousand AU can be extrapolated to scales of a
few hundred AU (Sch\"oier et al.\ 2004, J{\o}rgensen et al.\
2004b). Second, the drop-abundance profiles have been imaged directly
for a few species, in particular CO, H$_2$CO and N$_2$H$^+$. A
particularly good example is provided by L 483, shown in
Figure~\ref{l483} (J{\o}rgensen 2004).  C$^{18}$O is seen only in the
inner region and freezes out beyond 300 AU, at which point N$_2$H$^+$
becomes very prominent out to radii of about 8000 AU. At larger radii,
the density is too low for significant freeze-out, and also starts to
fall below the critical density for exciting lines of species like
N$_2$H$^+$. More generally, an accurate physical-chemical model of the
outer envelope is a prerequisite for interpreting interferometric data
of the inner envelope, since even if the line emission on larger
scales is resolved out, it can still affect the inferred abundances in
the inner region through optical depth effects (Sch\"oier et al.\
2004).

\section{Warm inner envelope}

Once the dust temperature reaches 90--100 K, even the most
strongly-bound ices like H$_2$O start to evaporate, resulting in a
`jump' in the gas-phase abundances of these molecules. In envelopes
around high-mass protostars, this temperature is reached at a radius
of more than 1000 AU, but it lies at less than 100 AU for the
low-luminosity objects considered here. Thus, the typical diameters of
these inner warm regions are less than 1$''$ and their emission is
severely diluted in the single-dish beams. Other effects such as
holes, cavities and disks also start to become important on these
scales. Nevetheless, if the abundance enhancements are sufficiently
large (typically more than a factor of 100), the effects may become
observable in the higher excitation lines.

An excellent example is provided by CH$_3$OH.  For high-mass sources,
jumps in its abundance of a factor of $\sim$100 have been derived by
simultaneously fitting the low- and high-excitation lines (e.g., van
der Tak et al.\ 2000). Since many of these lines can be obtained in a
single spectral setting (e.g., the $7_K-6_K$ setting at 338 GHz)
uncertainties due to calibration and different beam sizes are
minimized. For low-mass protostars like IRAS 16293--2422 and NGC 1333
IRAS2 similar abundance jumps from $\sim 10^{-9}$ to $\sim 10^{-7}$
around $T_{\rm ev}=90$~K are obtained (Maret et al.\ 2005, Maret, this
volume, J{\o}rgensen et al.\ 2005b, Sch\"oier et al.\ 2002). For other
sources like NGC 1333 IRAS 4B, however, the inferred $T_{\rm ev}$ is
much lower and the lines are much broader. Here the enhancement may be
due to the interaction of the outflow with the envelope which sputters
the ice mantles, such as observed in other outflows offset from the
central protostar (L1157: Bachiller \& P\'erez-Guti\'errez 1997; BHR71:
Garay et al.\ 1998; NGC 1333 IRAS2A: J{\o}rgensen et al. 2004c).
Which of these two mechanisms dominates the observed emission depends
on the relative filling factor of the thermal evaporation vs.\ shocked
zones in the observing beam.

A second example is provided by H$_2$CO. As an asymmetric rotor,
H$_2$CO has many lines originating from a wide range of energy levels
throughout the (sub)millimeter windows and is therefore an excellent
diagnostic probe (Mangum \& Wootten 1993, Ceccarelli et al.\
2003). However, even with so many lines, large abundance jumps in the
inner hot core region are difficult to establish unambiguously.  In
one interpretation, Maret et al.\ (2004) derived significant jumps up
to 3 orders of magnitude at $T_{\rm ev}=90$~K assuming an infalling
velocity structure and a fixed ortho/para-H$_2$CO ratio of 3. In an
alternative analysis, J{\o}rgensen et al.\ (2005b) find no evidence for
or against such abundance jumps using a turbulent velocity structure
and keeping the ortho/para-H$_2$CO ratio as a free parameter (inferred
to be typically 1.7). Both analyses agree well on the H$_2$CO
abundances in the outer envelope and their correlation with CO. Also,
both studies agree on a drop abundance profile with $T_{\rm ev}$=50~K
for IRAS 16293 -2422, the one source for which both interferometer data
and a large number of single-dish H$_2$CO lines are available
(Ceccarelli et al.\ 2000b, Sch\"oier et al.\ 2004).  These studies
illustrate the importance of future high-resolution multi-line interferometer
data to constrain the H$_2$CO abundance profile.


\section{Hot core chemistry: complex organics}

From the above analysis of the dust continuum and molecular line data,
it is clear that all low-mass protostars have inner warm regions where
the ices can, and often have, evaporated. This is one definition of
`hot cores'. High-mass hot cores are often characterized
observationally by a true forest of molecular lines down to the
confusion limit, including a multitude of features from complex
organics such as CH$_3$CN, CH$_3$OCH$_3$, HCOOCH$_3$ etc. In the
so-called `hot core chemistry', these molecules are thought to be the
second-generation products of gas-phase reactions with evaporated
molecules (e.g., Charnley et al.\ 1992). Until recently, it was
unclear whether low-mass protostars have a similar `hot core
chemistry'. 

\begin{figure}
\centering
\vspace{-1.5cm}
\includegraphics[width=0.8\linewidth,angle=-90]{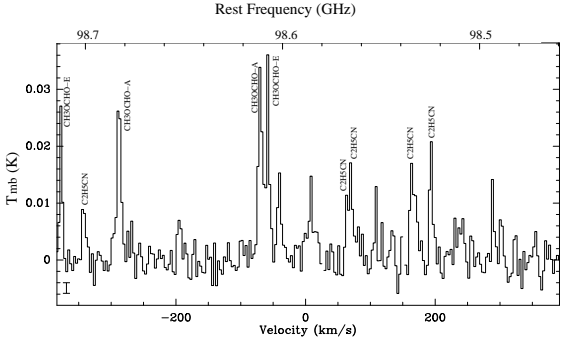}
\caption{Complex organic molecules observed toward IRAS 16293 -2422 with
the IRAM 30m telescope (Cazaux et al.\ 2003).\label{cazaux}}
\end{figure}

Deep single-dish integrations by Cazaux et al.\ (2003) have
beautifully demonstrated that low-mass objects like IRAS 16293 -2422
can have a similarly rich chemistry as high-mass protostars (see
Figure~\ref{cazaux}). Indeed, a recent JCMT survey by Caux et al.\ (in
prep.) of this object shows a remarkable line density. Moreover,
several of these molecules have been imaged with the SMA by Kuan et
al.\ (2004) and with the Plateau de Bure by Bottinelli et al.\
(2004b), showing chemical differentiation on arcsec scales between the
two sources in this protobinary object. A second example of a low-mass
protostar showing a rich line spectrum from complex organic species is
provided by NGC 1333 IRAS4A (Bottinelli et al.\ 2004a). Sch\"oier et
al.\ (2002) noted, however, that the timescales for crossing the hot
core region in low-mass objects are very short in a pure infall
scenario, only a few hundred yr, much shorter than the timescales of
$\sim 10^4-10^5$ yr needed for the hot core chemistry. Thus, a major
question is whether these species are actually first generation
molecules produced on the grains rather than the result of
high-temperature gas-phase chemistry. Alternatively, some mechanism
may have slowed down the infall allowing sufficient time for the hot
core chemistry to proceed.

\section{Conclusions and future}

A quantitative framework to constrain the physical and chemical
structure of protostellar envelopes on scales of 300--10000 AU has
been established and applied to a single-dish survey of a set of 18
low-mass objects.  The temperature and power-law density structure
derived from the single dish dust continuum data is found to work well
down to scales of a few hundred AU, although departures from spherical
symmetry and disks start to become important these scales.  In the
outer envelope, the chemistry is controlled by freeze-out of CO, which
also affects the abundances of other species like HCO$^+$ and
N$_2$H$^+$.  The multi-line data for most species are well fitted with a
`drop' abundance profile where the molecules are heavily depleted onto
grains in an intermediate cold zone.  Analyses of line profiles to
derive dynamical processes such as infall should take into account the
fact that the molecules may not be present throughout the entire
envelope.

In the inner envelope, the chemistry is controlled by the evaporation
of all ices, which can be released from the grains either by passive
heating or by sputtering in shocks. Complex organic molecules
characteristic of a hot-core chemistry have been found in at least two
sources, and interferometer data reveal small scale chemical
gradients. The origin of these complex molecules, in particular
whether they are first- or second-generation species, is still unclear.

There are a number of obvious future observing programs for Herschel
and ALMA.  Herschel will be unique in its ability to observe a
multitude of gas-phase water lines. Since H$_2$O is the dominant ice
species in the outer envelope, its freeze-out and evaporation will be
just as important in controlling the chemistry as that of CO. In the
innermost region and in shocks, high-temperature chemical reactions
can drive all of the oxygen into water, enhancing its abundance even
further. ISO-LWS has given a first glimpse of the rich science
associated with H$_2$O observation of low-mass Class 0 objects (e.g.,
Nisini et al.\ 2002, Maret et al.\ 2002), but Herschel will improve
these data by orders of magnitude in sensitivity, spatial and spectral
resolution.  ALMA will have unparalleled sensitivity to image many of
the molecules discussed here at spatial scales down to tens of AU,
revealing the chemical gradients in both the outer and the inner
envelope directly, and showing their relation with any circumstellar
disk and outflow material.

\section*{Acknowledgments}

Research in astrochemistry in Leiden is supported by a Spinoza grant
from NWO and by the Netherlands Research School for Astronomy (NOVA).


\end{document}